\documentstyle[aps,epsf,twocolumn]{revtex}
\begin{document}
\draft
\flushbottom
\twocolumn[
\hsize\textwidth\columnwidth\hsize\csname @twocolumnfalse\endcsname
\title{Open Luttinger liquids}
\author{A. H. Castro Neto$^{1}$, C. de C. Chamon$^{2}$,  and C. Nayak$^{3}$}

\address{
$^{1}$ Dept. of Physics, University of California, Riverside, CA, 92521\\
$^{2}$ Dept. of Physics, University of Illinois at Urbana-Champaign,
Urbana, IL, 61801\\
$^{3}$ Institute for Theoretical Physics,
University of California,
Santa Barbara, CA, 93106}
\date{\today}
\maketitle
\tightenlines
\widetext
\advance\leftskip by 57pt
\advance\rightskip by 57pt

\begin{abstract}
We study the problem of Luttinger liquids interacting with an active
environment. We are particularly interested in how dissipation affects
the response and correlation functions of non-isolated Luttinger
liquids. We show that the exchange of particles, energy, and momentum
lead to changes in the exponents characterizing the various
correlations functions. We discuss the importance of the zero mode
physics in this context.
\end{abstract}
\vspace{1mm}

]
\narrowtext
\tightenlines

When a single quantum mechanical degree of freedom, $X$, is the
cynosure in a system with many degrees of freedom, it is useful to
regard the rest as a bath which damps the dynamics of $X$ by
dissipating its energy \cite{Caldeira-Leggett} and, potentially, by
destroying its phase coherence \cite{leggett}.  Resistively shunted
Josephson junctions and heavy particles in a metal are examples of
physical systems which can be modelled by such a dissipative quantum
mechanics \cite{dissipation}.  It is very natural to wonder whether a
similar treatment is applicable to a subsystem of reduced
dimensionality which is spatially embedded in or at the boundary of a
larger system which plays the role of the bath. Precisely this
scenario appears to be realized in a number of different contexts.

Recent experiments in La$_{1.6-x}$Nd$_{0.4}$Sr$_x$CuO$_4$
and related compounds at $x=1/8$ indicate a realization of
a `striped' phase in which
Mott insulating antiferromagnetic regions are separated by anti-phase
domain wall boundaries \cite{tranquada}.  These domain walls, unlike
those of La$_2$NiO$_{4+x}$, are not empty \cite{tranquada}.  They have
a filling fraction of $1/4$, or one electron for every two sites, so
if the domain walls are treated as a system of quasi-one-dimensional
wires running through the CuO$_2$ plane, they should be metallic -- in
fact Luttinger liquid -- wires. However, the stripes are not simply
Luttinger liquids because they interact with the antiferromagnetic
spin fluctuations in the higher-dimensional Mott insulating regions
\cite{propaganda}, which can have several possible effects,
including superconductivity \cite{steve}.

Another example might be found in the quantum Hall regime \cite{hall}.
Recall that the edge excitations of a fractional quantum Hall state
ordinarily form a chiral Luttinger liquid \cite{chiral}.  However,
these Luttinger liquids are no longer isolated
when there is no gap in
the bulk, as is the case at most even-denominator filling fractions
\cite{even} and in quantum Hall ferromagnets, which have gapless spin
wave excitations in the limit of vanishing Zeeman energy
\cite{hallferro}. Such effects of the bulk on the edge
excitations could be detected in tunneling experiments\cite{Chang}.

One might worry that it does not always make sense to make this
distinction between the subsystem and the bath when the subsystem has
infinitely many degrees of freedom and interactions between the
subsystem and the rest of the system are not small.  The excitations
of the subsystem mix with those of the bath and the two cannot really
be separated.  Here, however, we consider the situation in which the
subsystem is a spatially well-defined entity, as in the examples given
above. In these cases, this division is particularly natural because
certain types of transport measurements couple exclusively to the
subsystem.

The preceding considerations serve as motivation.  In this paper, we
will be primarily concerned with setting up and solving a simple model
of a Luttinger liquid interacting with a bath. First, we set up the
model of a scalar field $\theta$ interacting with a bath.
Integrating out the degrees of freedom of the bath induces
non-local effective interactions in the scalar field. We
focus on two novel aspects of the resulting theory: the non-local
interactions which can be controlled by manipulating the bath,
e.g. by varying the temperature; and the dissipation of
the zero mode of the scalar field, as a result of which the
conserved quantity (such as charge or spin) can
leak into the bath along the whole length
of the system (rather than just through external leads).
We show how the combined effects
modify the correlation functions of the
operator $e^{i\theta}$. Finally, we discuss the implications for
one-dimensional electron systems. We will only touch lightly on the
applicability of this model to specific systems; a more thorough
discussion, including comparisons with experimental data, will be
reserved for a later publication.

The model we consider is a scalar field interacting with a bath of
oscillators with some specifiable spectrum. Our primary concern is
interacting one-dimensional electron systems, for which the scalar
field is the bosonized representation of the electronic degrees of
freedom. This representation is obtained by writing the fermion
operator \cite{boso}
\begin{equation}
\psi(x,t) \propto \sum_{n\ {\rm odd}}
\ e^{i n \sqrt{\pi}\left(\theta(x,t)+k_Fx\right)}
\ e^{i\sqrt{\pi}\phi(x,t)}
\end{equation}
in terms of non-independent boson fields $\phi$ and $\theta$, which
satisfy the commutation relation $[\phi(x),\theta(x')]=-i\ {\rm
sgn}(x-x')$. The action associated with these scalar fields and their
interaction with the bath (we use units such that $\hbar = k_B =1$) is
given by
\begin{equation}
S = S_{{\cal L}} + S_{B} + S_{{\cal L}B}
\label{action1}
\end{equation}
where $S_{{\cal L}}$ is the action of the scalar fields (which
describes a Luttinger liquid),
\begin{equation}
{S_{{\cal L}}} =
\frac{1}{2 g}\int_{x,t} \,{\left({\partial_\mu}\theta\right)^2}
=
\frac{g}{2}\int_{x,t} \,{\left({\partial_\mu}\phi\right)^2},
\end{equation}
where $g$ is the Luttinger parameter that describes the forward
scattering interactions between the fermions ($g<1$, $g=1$, $g>1$
for repulsively interacting, free, and attractively interacting
fermions, respectively) and the velocity of propagation is set to
one ($\mu = 0,1$ for the time and the one spatial coordinate). The
fields $\phi$ and $\theta$ provide two dual descriptions of the
Luttinger liquid. Alternatively, we may choose to work in a chiral
basis, where we write $\theta={\chi_R}+{\chi_L}$ and
$\phi=g^{-1}({\chi_R}-{\chi_L})$. As a consequence of the equations
of motion, the field $\phi$ can be expanded as
\begin{eqnarray}
\chi_{R,L}(x,t) &=& \sqrt{\frac{g}{4 \pi}} \left\{
X_{R,L} + {P_{R,L}} \frac{2 \pi}{\ell} (t \mp x) \right.
\nonumber
\\
&+& \left. \sum_{n>0}{i\over n}\left[
a_{R,L;n}\ e^{-i{2\pi \over \ell}n(t\mp x)}-
c. c.
\right] \right\}\
\label{decompose}
\end{eqnarray}
where $\ell$ is the size of the system.  The canonical
quantization condition leads to the following
commutation relations
$[a_{R,L;n},a^\dagger_{R,L;m}]=n\delta_{n,m}$,
$[X_{R,L},P_{R,L}]=i$.
In the absence of the bath,
$X_{R,L},{P_{R,L}},{a_{R,L;n}},{a_{R,L;n}^\dagger}$ are
time-independent. $P_{R,L}$ are right-moving and left-moving fermion
numbers; $X_{R,L}$ are the conjugate phases.  (In the language of
conformal field theory, $P = {P_R} + {P_L}$ and $W = {P_R} - {P_L}$
are, respectively, the momentum and winding number of the scalar
field.)  When the coupling to the bath is turned on, these operators
acquire time-dependence. Finally, the Hamiltonian associated with the
decomposition reads,
\begin{eqnarray}
H_{{\cal L}} = \frac{2 \pi}{\ell} \left\{\frac{P_{R}^2 + P_{L}^2}{2} +
\sum_{n>0} \left[ a_{R,n}^\dagger a_{R,n}+
a_{L,n}^\dagger a_{L,n} \right] \right\}\, .
\label{freehamil}
\end{eqnarray}

The dissipation arises from the interaction of the one dimensional
system with the environment (the higher dimensional embedding system).
This coupling will depend on the nature of the bath, and which degrees
of freedom of the one dimensional system are sensitive to the
fluctuations of the bath. More precisely, the environment could be a
bath of photons, phonons, spin waves, or most generally, any set of
collective excitations, which share the common feature of being
described as a collection of harmonic oscillators. It is the physical
nature of the bath that determines which sort of densities in the one
dimensional system are affected by the environment, for example charge
currents for coupling to a photon bath, and spin currents for coupling
to a magnetic bath.

Here we will work with interactions which can be written in
two ways: {\it (A)} In terms of
either a coupling between a vector current and a vector field,
\begin{equation}
S_{{\cal L}B}=\int_{x,t}  j_\mu(t,x)\ A^\mu(t,x,{\bf x_\perp}=0),
\label{LBintactionvector}
\end{equation}
where $j^\mu=\epsilon^{\mu\nu}\partial_\nu\theta/\sqrt{\pi}$ is the current
obtained from the scalar field $\theta$, and $A^\mu$ is a quantized
harmonic field, and can describe
any vector-like collective excitation for the bath;
{\it (B)} In terms of density and a scalar field,
\begin{equation}
S_{{\cal L}B}=\int_{x,t} \partial_x\theta\ \sum_\alpha \lambda_\alpha\
\Phi_\alpha(t,x,{\bf x_\perp}=0),
\label{LBintactionscalar}
\end{equation}
where $\alpha$ labels different modes, which couple to the density
fluctuations $\delta \rho=\partial_x\theta/\sqrt{\pi}$ with different
weights $\lambda_\alpha$. In both cases the bath Hamiltonian is simply
that of a set of harmonic oscillators with characteristic
frequencies ${\omega_{{\bf k},\alpha}}$ which contain the
information about the bath. Notice that the bath have extra degrees
of freedom as compared to the 1-D system, which can be labelled by the
perpendicular momenta ${{\bf k_\perp}}$ (the coupling to the 1-D
system takes place at ${{\bf x_\perp}=0}$).

This model is the natural generalization of that of a point particle
interacting with a dissipative environment. The bath Hamiltonian is
identical and the bilinear coupling of the bath to the particle's
position has become a similar coupling to the field, $\theta$ or
$\phi$. The point particle is replaced by a free scalar field, which
can be viewed, from a first-quantized standpoint, as a collection of
non-interacting point particles. As in the point particle case, we can
integrate out the bath to derive a non-local but {\it quadratic}
effective action. This quadratic effective action, which we analyze
below, is one of the main results of this paper.

Now, in contrast to the single particle, the field has many degrees of
freedom labelled by the longitudinal momentum $k$.  Each $k$ mode of
$\theta$ or $\phi$ is dissipated independently, so we can generalize
the results for a single particle in a dissipative environment in a
simple way. Integrating out the harmonic bath ({\it i.e.}, the fields
$A^\mu$ or $\Phi_\alpha$), we are left with the following (Euclidean)
action for the $\theta$ field:
\begin{equation}
S = \frac{1}{2} \int_{\omega,k}
\left[\frac{1}{g} \left(\omega^2+k^2\right)+J(k,\omega)\right]
|\theta(k,\omega)|^2\ .
\end{equation}
For each $k$ mode, the properties of the bath can be summarized by a
spectral function $J(k,\omega)$. Let us then turn to the analysis of
the consequences of different forms for $J(k,\omega)$. Consider
spectral functions that, for small $k$ and $\omega$ behave as
$J(k,\omega)\propto|k|^\alpha|\omega|^\beta$. We have the following
possibilities:
{\it (i)} $\alpha+\beta>2$ - the effect of the bath on the long time
or long distance physics is irrelevant. However, there may still be
effects due to total charge dissipation if the zero mode couples to
the bath. This will be considered later.
{\it (ii)} $\alpha+\beta=2$ - this is the marginal case.  The effect
of $J(k,\omega)$ is to renormalize the Luttinger parameter $g$ in
front of the logarithmic correlation function for
$\theta$\cite{t-x}. This occurs, for instance, in the
case of a Luttinger liquid interacting with phonons,
analyzed in \cite{martin}.
The usual case of short range interactions, which
shifts $g$, can be obtained by integrating out a bath of
massive photons; the effect of the bath, as presented in this paper,
is just a more general way to generate and think about
interactions. We call attention to the fact that, in this picture,
$g$ arises from the
bath and therefore may acquire dependence on an external parameter
({\it e.g.} temperature) affecting such bath.
{\it (iii)} $\alpha+\beta<2$ - in this case the bath leads to an
energy gap in the spectrum of $\theta$. One may worry that this will
lead to inconsistences if we look at correlations of $\psi\propto
e^{i\theta}$, since the gap implies that $\lim_{t\to
\infty}\langle\theta(t,0)\theta(0,0)\rangle=0$, and so $\psi$ acquires
long range order. Kogut and Susskind \cite{Kogut&Susskind} have
studied this problem in the context of quark confinement. They have
termed the vanishing of
$\lim_{t\to\infty}\langle\theta(t,0)\theta(0,0)\rangle=0$ the ``seizing''
of the vacuum: the interactions generated by the $J(k,\omega)$ are
{\it long ranged}, which leads to {\it true ordering} in our 1+1
dimensional system. Even though there is a broken symmetry (shifts in
$\theta$), there are no Goldstone bosons associated with this broken
symmetry; the infinite range interactions imply a finite energy cost
even for long wavelength fluctuations. This case is especially
interesting. The connected part of the electron Green's function
$\langle\psi(t,0)\psi^\dagger(0,0)\rangle_c$ vanishes except at $t=0$. In
other words, electrons are confined, so that this corresponds to a
novel type of insulator phase where conduction is supressed because
free carriers are confined. The gap to excitations in the charged
sector is proportional to the coupling to the bath.

As we mentioned previously, the effect of the bath is a more general
way to think about interactions. Indeed, the bath can generate
interactions which can be tuned via an external parameter, such as
temperature or magnetic field, in such a way that one may
switch among the three cases {\it
(i),(ii),(iii)} above. Consider, for example, a bath of bosonic
exitations in 2+1 dimensions, generating a $J(k,\omega)\propto
k^2/\sqrt{k^2+m^2}$, where $m$ is a mass gap. This mass measures a
correlation length in the bath, $m\sim\xi^{-1}$, which can be, say,
temperature dependent. Clearly, when $\xi\to \infty$, we are in case
{\it (iii)}. For any finite $\xi$, $J(k,\omega)\propto \xi k^2$, we
are in case {\it (ii)} and the Luttinger parameter inherites the
temperature dependence of the bath through $\xi$.
For the cases {\it (i)} and {\it (ii)} corrections due to dissipation of
the charge (zero mode) could be qualitatively important. Let us now
turn to this problem.

If the subsystem exchanges charge (or spin) with the bath,
in addition to energy
and momentum, we must include the effects of the dissipation of the total
charge of the subsystem. Here, we have in mind the possibility
of charge being exchanged with the bath along the length
of the system rather than a coupling to leads or
point-contacts.
To do so, we take proper care of the
zero mode or total charge mode. We then allow for spectral
functions that have a finite weight at $k=0$:
\begin{equation}
J(k,\omega)=J_0(\omega)\ \delta(k) + J_{reg}(k,\omega)\ .
\label{zmdisscoupl}
\end{equation}

Several comments are in order. There are alternative ways
of incorporating charge non-conservation in our model, such
as coupling the bath directly to the operator $e^{i\theta}$.
We have chosen a simple (and tractable) way of incoporating
this phenomenon. As we show below, even the simple coupling
(\ref{zmdisscoupl}) has interesting consequences. It is tempting
to speculate on the possible similarity between the behavior resulting
from different charge-dissipative couplings \cite{lev}.
A second important point regarding (\ref{zmdisscoupl})
is the $\delta$-function strength of the dissipation
of the zero mode (which we justify in another way below),
which is necessary to give a non-vanishing
effect in the infinite-size limit. This, in turn, guarantees
that the zero-mode dynamics has a non-vanishing effect on the
Green function whereas, ordinarily, a single $k$ mode with unit normalization
would have vanishing contribution.

The zero mode of the scalar boson field $\theta$ can be read from the
mode decomposition Eq. (\ref{decompose}), $
\theta_0(x,t) = \sqrt{\frac{g}{\pi}}
\left\{ X(t)+\pi W x/\ell\right\}\ ,$
where $X=(X_R+X_L)/2$, $P=P_R+P_L$, $W=P_L-P_R$, and in the last
step we moved to the Heisenberg representation of $X,P$ and $W$
(notice that $X(t)=X+ \pi P t/\ell$). Then
$\theta(k=0,t)=\sqrt{g/\pi}\, \ell\ X(t)$.
We can also find a similar representation in
terms of $\Theta=(X_R-X_L)/2$ which is the conjugate of the winding
number $W$. Dissipation of the total current of the system enters
through $X(t)$, whereas dissipation of the total charge enters through
$\Theta(t)$. These two cases can be treated similarly, so for
simplicity we consider below dissipation in just one, say $X$.
Neglecting $J_{reg}(k,\omega)$ for the moment, we find an action for
$X$ of the form
\begin{equation}
{S_X}=\frac{1}{2}\int_{\omega}
\left(\frac{\ell}{\pi}\ \omega^2+ \frac{g \ell^2}{2 \pi^2} J_0(\omega)\right)
|X(\omega)|^2 \ .
\label{zmaction}
\end{equation}
Notice that this is simply the dissipative action for a single
particle of mass $M=\ell/\pi$, which here is our
zero mode.
The case of greatest interest, ohmic dissipation,
$ J_0(\omega) \propto |\omega|$
has been studied in great detail in the literature \cite{dissipation}.

Since the dissipation is related to the collective mode of the bosonic
field one expects to obtain Lagrangian (\ref{zmaction}) from a
first-quantized viewpoint by treating the center of mass,
$R_{CM}(t) = \sum_i x_i(t)/N$, of the
bosons as the dissipative particle ($N$ is the boson number).
For ohmic dissipation the
effective action (after the trace over the modes) reads
\cite{Caldeira-Leggett}
\begin{eqnarray}
S_d =\frac{\eta_P}{2 \pi} \,\int\,dt\,dt'\, \frac{R_{CM}(t)
R_{CM}(t')}{(t-t')^2}
\end{eqnarray}
where $\eta_P$ is the friction coefficient. Using the definition
of the center of mass and writing the action in terms of the fermion densities,
$\rho(x,t) =\sum_i \delta\left(x-x_i(t)\right) =
\rho_0+{\partial_x}\theta(x,t)/\sqrt{\pi}$
where $\rho_0 = N/\ell$ is the average density, we find,
\begin{eqnarray}
J_0(\omega) = \frac{\eta_P}{\ell^2 \rho_0^2} |\omega|
\label{j0}
\end{eqnarray}
which gives the friction coefficient for the bosonic collective mode,
$\eta_P/\rho_0^2$.

The damping of the zero mode (\ref{j0}) leads to the dissipation into
the bath of the conserved charge associated with
the scalar field -- either the charge or the spin of the bosonized
electrons.  This has profound consequences for the correlation
functions of the operator $e^{i\theta}$.
The $|\omega|$ dissipation term dominates the effective action
(\ref{zmaction}) for low energies. Thus, for long times ($ 1 >> t/\ell
>> 2 \pi \rho_0^2/(g \eta_P)$),
$\langle X(t) X(0) \rangle_c \approx \pi\rho_0^2/(g \eta_P) \,
\ln[g \eta_P t/(2\pi \rho_0^2 \ell)]$.  In this case the
correlation functions such as $\langle {e^{i\theta(t,0)}}
{e^{-i\theta(0,0)}} \rangle$ are modified from their undamped forms by a
factor of $t^{-\rho_0^2/ \eta_P}$. Similarly, dissipation
of charge would modify $\langle {e^{i\phi(t,0)}} {e^{-i\phi(0,0)}}
\rangle$ by a factor $t^{-\rho_0^2/ \eta_W}$, where $\eta_W$ is the
friction coefficient for the total charge. These corrections
explicitly alter the electron propagator in the 1-D system:
\begin{eqnarray}
G(x,t) &=& \langle \psi(x,t) \psi^\dagger(0,0) \rangle
\nonumber
\\
&\approx&
\frac{G_0(x,t)}
{[\frac{g \eta_P t}{2\pi \rho_0^2 \ell}]^{\pi \rho_0^2/\eta_P}
[\frac{\eta_W t}{2\pi g\rho_0^2 \ell}]^{\pi \rho_0^2/\eta_W}}
\label{newgreen}
\end{eqnarray}
where $G_0(x,t)$ is the correlation function in the absence of charge
dissipation. Our conclusion is therefore that for long times the
electron correlation function decays like $t^{-\gamma}$, where
$\gamma=(g+g^{-1})/2 + \pi \rho_0^2 (\eta_P^{-1}+\eta_W^{-1})$.
This correlation function decays faster than the usual Luttinger
liquid correlation function. In many examples of interest,
$\eta_{P,W}$ will be temperature-dependent, and the Green function
will have unusual temperature dependence.

This is precisely the dissipation of charge that we mentioned
earlier. As we will discuss further, the damping changes the Luttinger
liquid exponents for tunneling at a point contact. Naively, the
physics behind $\eta$ is quite different from the physics relating to
the Luttinger liquid parameter, $g$. The former reflects the diffusion
of charge away from the Luttinger liquid (in a position independent
way); the latter reflects the reduced density of states at the Fermi
points due to repulsive interactions. However, for point contact
tunneling, they have a similar effect because the Luttinger liquid
plays the role of a bath and, hence, $\frac{1}{2}(g+g^{-1})$ and
$\pi\rho_0^2 (\eta_P^{-1}+\eta_W^{-1})$ play similar roles.

In an electron system which is not spin-polarized, there is a
non-trivial spin sector. The charge sector is similar to that of
spinless (i.e. spin-polarized) electrons; if it is coupled to a bath,
the charge dynamics is determined by the previous results.
If, on the other hand, the spin sector is coupled to a bath
while the charge sector is undamped, both charge and spin dynamics can
be anomalous.  The latter is not surprising; the former occurs because
electron tunneling necessarily involves the recombination of spinon
and holons.
In the abelian bosonization scheme, both the charge and spin sectors
are described by free scalar fields with Luttinger liquid parameters
${g_c}$ and ${g_s} = 2$. The electron operator,
${e^{i\phi_c}}\,{e^{i\phi_s}}$ is a dimension
$(1/g_c+1/g_s)/2$ operator.  In
the absence of dissipation, this determines the temperature dependence
of tunneling at a point contact, $G \sim {T^{2(1-1/{g_c}-1/{g_s})}}$.
In the presence of dissipation, the dimension of the electron
tunneling operator is shifted by $\pi\rho_0^2\eta^{-1}_{s,c}$
through the shift in the spinon operator, so $G \sim
{T^{2(1-1/{g_c}-1/{g_s}-\pi \rho_0^2\eta^{-1}_{s,c})}}$, in
complete analogy with the previous discussion.

In summary, we have, in this paper, considered the
simplest possible `toy model' of
a Luttinger liquid interacting with a dissipative bath. Remarkably, we
find interesting behavior even in this model such as modified Luttinger
liquid exponents which depend on the spectrum of the bath.
Possible consequences for the QHE and cuprates
will be considered elsewhere.

We thank A. Balatsky, G. Castilla, E. Fradkin, D. E. Freed,
M. R. Geller, A.W.W. Ludwig, S. Kivelson, C. Mudry , S. H. Simon, and
X.-G. Wen for
discussions. The authors acknowledge the Aspen Center for Physics for
the hospitality during the initial stages of this work. This work was
supported in part by the National Science Foundation through grants
NSF-DMR-94-24511, NSF-DMR-89-20538 and NSF-ITP-97-029.



\begin{references}
\bibitem{t-x} Even when $\alpha\ne\beta$,
the spatial and temporal dependence of the correlation functions are
affected in the same way so long as $\alpha+\beta =2$.
\end{references}
\end{document}